\documentclass[iop,apj,useAMS,usenatbib]{emulateapj-rtx4}
\usepackage{graphicx}
\usepackage{adjustbox,lipsum,natbib,epstopdf,color,amsmath,epsfig,url}
\usepackage{float}
\usepackage[caption = false]{subfig}
\usepackage{hyperref}
\hypersetup{colorlinks=true, citecolor=blue, filecolor=magenta,
  urlcolor=cyan, linkcolor=blue}

\begin{document}
\author{Farhanul Hasan\altaffilmark{1}, Hooshang Nayyeri\altaffilmark{2}, Asantha Cooray\altaffilmark{2}}
\altaffiltext{$\star$}{{\it Herschel} is an ESA space observatory with
  science instruments provided by European-led Principal Investigator
  consortia and with important participation from NASA.}
\affil{$^1$ Reed College, Portland, OR 97202}
\affil{$^2$ Department of Physics and Astronomy, University of California, Irvine, CA 92697}

\title{Far-infrared and Nebular Star-formation Rate of Dusty Star Forming Galaxies from  {\it Herschel}$^*$ and 3D-HST at $\lowercase{z}\sim1$}

\begin{abstract}
We present results of a multi-band Spectral Energy Distribution (SED) and nebular emission line analysis of a sample of 1147 spectroscopically identified dusty star-forming galaxies at $0.49<z<2.24$ from {\it Herschel}/SPIRE and {\it HST}/WFC3 grism observations in the five CANDELS fields: AEGIS, GOODS-N, GOODS-S, COSMOS, and UDS. We use the spectroscopic redshifts measured from nebular lines to construct the SEDs of galaxies from the optical to the infrared using {\it HST} and {\it Herschel} photometry. We further utilize the 3D-HST grism H$\alpha$ line flux measurements to measure the nebular star-formation rates after correcting for attenuation. We compare this with direct observations of the SFR measurements in the far-infrared from {\it Herschel}. Observation of the infrared excess (IRX) in this sample as a function of the UV spectral slope reveals that these DSFGs deviate toward toward bluer colors, thus sitting well above the expected relation for normal star-forming galaxies. The high-$z$ dusty galaxies have a stellar mass distribution that is skewed towards larger masses, with $\rm M_{med}=2.6\times10^{10}\,M_{\odot}$. However this population has star-formation rates consistent with the most massive tail of the main sequence, showing that these are both the most massive and the most star-forming galaxies during the peak epoch of formation.
\end{abstract}


\section{Introduction}

Galaxies assemble their stellar mass by converting gas into stars within cold giant molecular clouds \citep{Shu1987, Lada2010, Murray2010, Schneider2015}. The rate at which the new stars are formed (the so-called star-formation rate or SFR; \citealp{Schmidt1959, Kennicutt1998}) is one of the key factors in determining the evolution of a galaxy \citep{Hopkins2011, Muzzin2013, Lilly2013, Gonzalez2014, Furlong2015}. Dusty Star-forming Galaxies (DSFGs; for a recent review see \citealp{Casey2014}) are among the most star-forming galaxies in the universe \citep{Riechers2014, Nayyeri2017a}. These systems are bright in the far-infrared where the UV light emitted by young stars are re-emitted by dust in longer wavelengths \citep{Reddy2012, Casey2014}. 

{\it Herschel} Space Observatory \citep{Pilbratt2010} revolutionized studies of the far-infrared universe by allowing direct observations of dust in the local and distant universe \citep{Magnelli2012, Elbaz2011, Gruppioni2013, Schreiber2015}, characterization of rate of major mergers as a function of infrared luminosity and role of merger driven starburst activity \citep{Rosario2012, Schreiber2015} and studies of the background far-infrared light \citep{Berta2011, Cooray2012, Zemcov2014}. {\it Herschel} observations of individual sources provided unique data sets to construct the far-infrared spectrum of dusty galaxies for the first time where previous such measurements were limited to indirect estimates, such as from the mid-IR \citep{Reddy2010}, with inherent uncertainties associated with the conversion and the dust heating \citep{Elbaz2011, Reddy2010}.

One of the main goals achieved with {\it Herschel} is probing the dust obscured star-formation activity in nearby and distant galaxies \citep{Elbaz2011, Reddy2012, Overzier11, Lee2012, Mancuso16}. Direct observations of the dust obscured systems in the far-infrared is used to constrain the universal infrared star-formation rate density budget and to investigate the relative importance of different populations of IR luminous galaxies to this as a function of redshift \citep{Casey2014}. Measurements of the star-formation rates from infrared observations (as a direct probe of obscured SFR; \citealp{Elbaz2011, Whitaker2012}) are calibrated from infrared luminosities (assuming an initial mass function; \citealp{Shu1987, Kennicutt1998}) and could be directly compared to other SFR indicators \citep{Reddy2010, Hao2011, Reddy2012, Burgarella2013}. This comparison yields valuable information about not only the hidden modes of star-formation but also combined with the different diagnostics provide insight into the burstiness of the star-formation activity and hence time scales of star-formation and is a good diagnostic of the star-formation history of distant galaxies \citep{Wuyts2011, Reddy2012, Madau2014}. 

Another main indicator of star-formation activity in galaxies is nebular emission lines \citep{Osterbrock1989} -- in particular, the Hydrogen Balmer line H$\alpha$ at 6563\AA. The recombination emission line H$\alpha$ is produced in ionized H\,II regions surrounding the most massive O stars and hence provides a direct probe of instantaneous and the bursty star-formation activity in galaxies at short time scales of $\sim10$\,Myr. On the other hand, underlying UV continuum in galaxies is produced by the less massive young stars and traces star-formation activity over longer average time scales of $\sim100$\,Myr. Tracing these different modes of star-formation is crucial in forming a complete picture of total star-formation in a galaxy \citep{Reddy2010}. In particular, the different SFR diagnostics are affected differently by dust, given the wavelength, with the UV SFR being affected the most and IR the least. Hence comparison of these different diagnostics provide indirect probes of the attenuation in galaxies. 

In this work, we use combined {\it Herschel} and 3D-HST \citep{Brammer2012} observations of the five Cosmic Assembly Near-Infrared Deep Extragalactic Legacy Survey (CANDELS; \citealp{Grogin2011, Koekemoer2011}) fields with optical and near-infrared observations from {\it HST}/ACS and WFC3 to construct a complete census of dusty star-forming galaxies at $z\sim1$. In particular, we use the {\it Herschel} far-infrared observations to measure the total far-infrared star-formation rates. We combine this with 3D-HST grism observations of the nebular lines and the {\it HST}/ACS and WFC3 photometry to get a full construction of the SED of DSFGs from optical to far-infrared and to study the different modes of star-formation from UV to FIR. Additionally, we use these observations to probe the different attenuations that drive different star-formation diagnostics.

This paper is organized as follows. In Section 2 we present the data used in this work, which includes observations by {\it Hubble} and {\it Spitzer} Space Telescopes and {\it Herschel} Space Observatory. We discuss our measurements of the physical parameters of the sources from multi-band SED fits and observed line fluxes in Section 3. The analysis of our results are presented and discussed in the context of dust attenuation in DSFGs in Section 4. We use a Chabrier initial mass function (IMF; \citealp{Chabrier2003}) throughout this paper. We further assume a standard cosmology with $H_0=70\,\text{kms}^{-1}\text{Mpc}^{-1}$, $\Omega_m=0.3$ and $\Omega_\Lambda=0.7$. All magnitudes are in the AB system where $\text{m}_{AB}=23.9-2.5\text{log}(f_{\nu}/1\mu \text{Jy})$ \citep{Oke1983} unless otherwise noted.

\section{Data} 

Our sample consists of {\it HST} grism observations of $z\sim 0.5-1.6$ sources from the 3D- HST survey\footnote{\url{ http://3dhst.research.yale.edu}} \citep{Skelton2014, Momcheva2016} in the five major fields observed by {\it HST} Wide Field Camera 3 as part of the Cosmic Assembly Near-infrared Deep Extragalactic Legacy Survey (CANDELS; \citealp{Grogin2011, Koekemoer2011}). Particularly, we take advantage of observations by the \textit{Herschel} Space Observatory as part of the {\it Herschel} Multi-tiered Extragalactic Survey (HerMES; \citealp{Oliver2012}). This data-set includes photometric observations from {\it Hubble} and {\it Spitzer} Space Telescopes and {\it Herschel} Space Observatory from the optical to the infrared bands and spectroscopic observations with {\it HST}/WFC3 grism. As such, galaxies in our sample have ample ancillary observations that are used to construct their Spectral Energy Distributions (SEDs) and to measure their physical properties. 

\subsection{{\it Herschel}/HerMES}

The {\it Herschel} Multi-tiered Extragalactic Survey ({\it HerMES}; \citealt{Oliver2012}), is a legacy survey program designed to study the evolution of galaxies in the distant universe and was the largest project on the European Space Agency's (ESA) {\it Herschel} Space Observatory (900 hours; \citealt{Pilbratt2010}). In this work we used the fourth data release of the HerMES public photometric catalogs available on HeDaM\footnote{\url{http://hedam.oamp.fr/HerMES/}}. This included observations at 250\,$\mu$m, 350\,$\mu$m and 500\,$\mu$m bands by the Spectral and Photometric Imaging REceiver (SPIRE; \citealp{Griffin2010}). We used the XID250 catalogs which provides SPIRE photometry for objects whose positions are taken from catalogs extracted from {\it Spitzer} MIPS 24\,$\mu$m maps \citep{Roseboom10, Roseboom12}. Furthermore, we used the COSMOS XID+ catalog which uses the XID+ tool \citep{Hurley17}. This catalog uses 24\,$\mu$m-detected sources from the MIPS 24\,$\mu$m catalog \citep{LF09} as a prior list for extracting SPIRE fluxes from the HerMES SPIRE maps \citep{Oliver2012}. The positional information determined from MIPS observations allows us to cross-match the {\it Herschel} observations with the near-infrared and optical data observed by HST as discussed below. MIPS Sky coordinates were matched with {\it HST} coordinates, allowing for a positional error of up to $2^{\prime\prime}$. About 5\% of the cross-matches were misidentifications that were later removed.

\begin{figure}[htbp]
\begin{center}
\includegraphics[trim=2cm 0cm 0cm 0cm, scale=0.45]{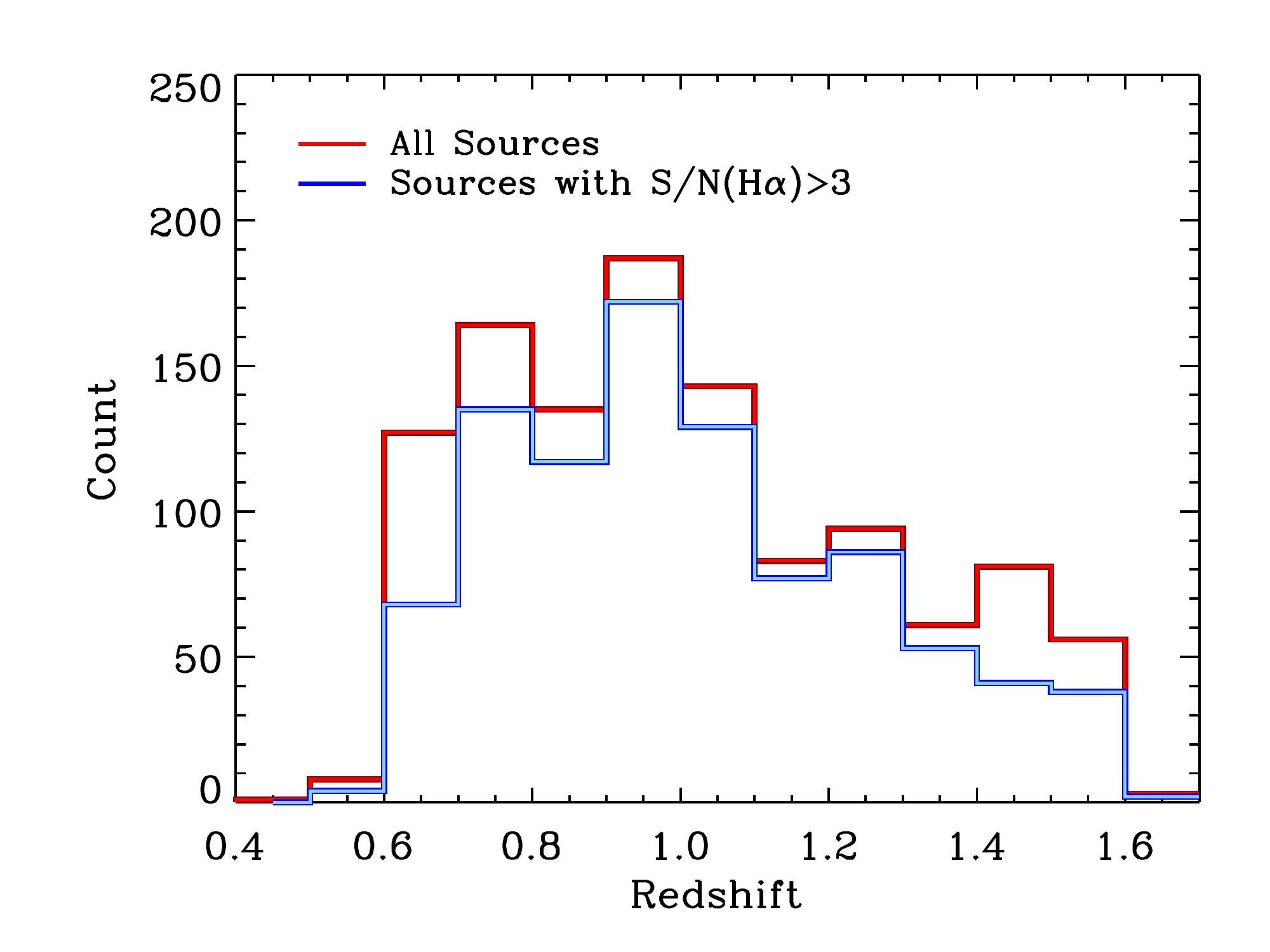}
\caption{The 3D-HST spectroscopic grism redshift distribution for the combined {\it HST}+{\it Herschel} sample of 1147 sources with reported H$\alpha$ used in this study. The sub-sample of sources with 3$\sigma$ detection of the H$\alpha$ line is shown in blue. A good fraction of the sources in our sample have reliable detection of the H$\alpha$ line.}
\label{redshift}
\end{center}
\end{figure}

\begin{table}
\begin{center}
\caption{Source match between {\it Herschel} and {\it HST}}
\begin{tabular}{lccccc}
\hline
Field   & All $^a$ & All H$\alpha$$^b$ & All H$\beta$$^b$   & H$\alpha$$^c$ & H$\alpha$ and H$\beta$$^d$  \\ 
\hline
\hline
AEGIS   & 348           & 55   & 17   & 42 & 0             \\
COSMOS  & 1369                   & 338  & 50  & 282  & 7            \\
GOODS-N & 2098                   & 323  & 63  & 274 & 11           \\
GOODS-S & 1911                   & 379 & 82 & 297 & 8          \\
UDS     & 292                    & 52   & 11   & 30 & 0          \\ 
\hline
Total   & 6018                   & 1147 & 223 & 925 & 26      \\
\hline  
\end{tabular}
\label{matches}

\end{center}
\footnotesize{
$^a$: All sources matched between the {\it Herschel} and 3D-HST catalogs with a $2^{\prime\prime}$ radius. $^b$: All matched sources with H$\alpha$ or H$\beta$ reported. $^c$: Matched sources with at least 3$\sigma$ detection of H$\alpha$. $^d$: Matched sources with 3$\sigma$ detection in H$\alpha$ and H$\beta$.}

\end{table}

\begin{figure}[htbp]
\centering
\includegraphics[trim=2cm 0cm 0cm 0cm, scale=0.45]{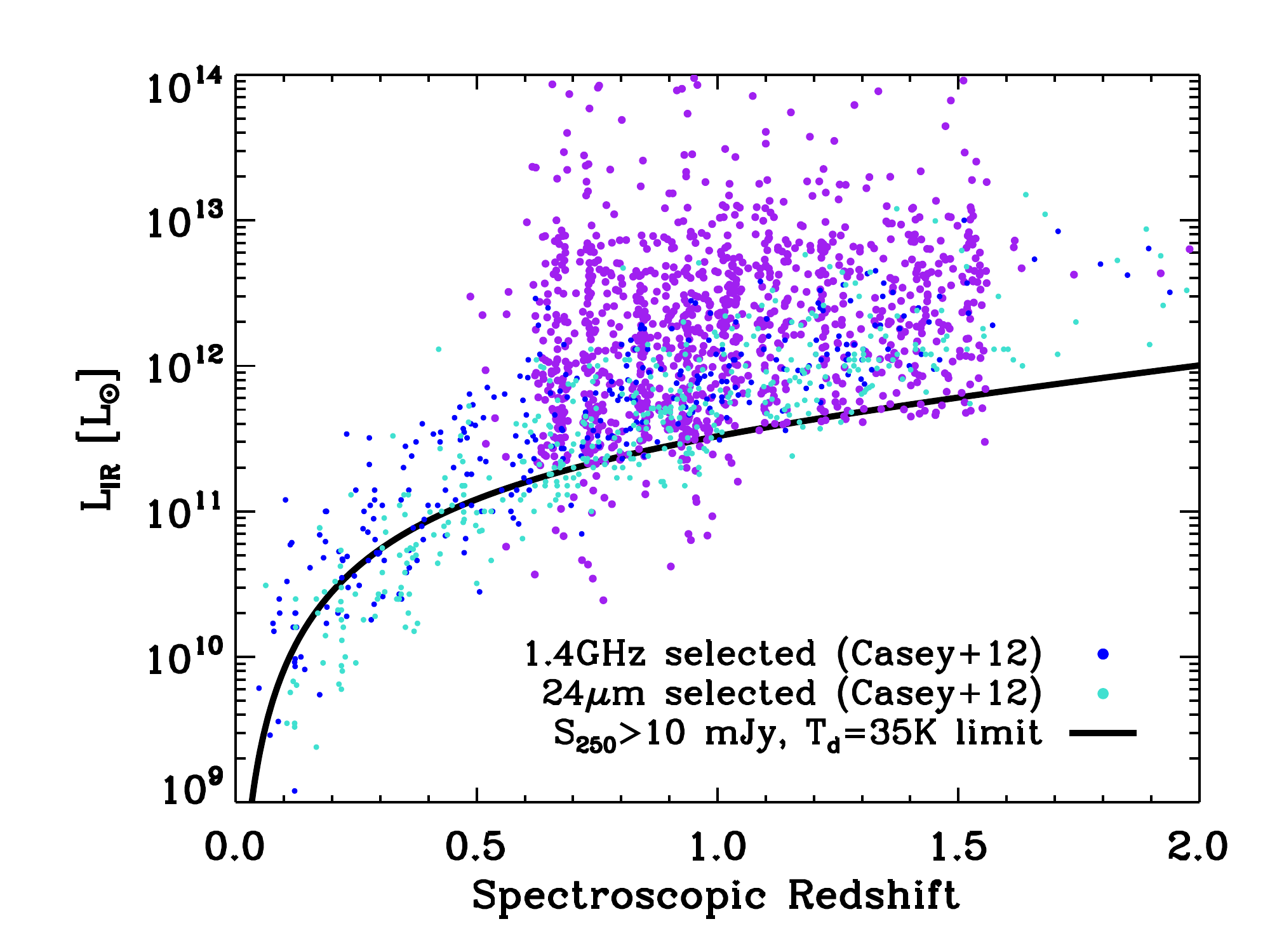}
\caption{Far-infrared luminosities ($\rm L_{IR}$; rest-frame $\rm 8-1000\,\mu m$) against spectroscopic redshift for our sample of {\it Herschel} detected DSFGs. The black solid line denotes the {\it Herschel} detection limit of $\rm S_{250} > 10 mJy$ for a modified black-body with $\rm T_{d} =35 K$ \citep{Casey2012a}. Our galaxies are well detected by {\it Herschel}, sitting well above the detection limit across all redshifts. For comparison, we are showing infraed luminosity distributions of samples of 24\,$\mu$m and 1.4\,GHz selected sources from \citet{Casey2012b}.}
\label{fir-z}
\end{figure}

\subsection{3D-HST Grism Observations}

The 3D-HST near-infrared grism spectroscopic survey \citep{Brammer2012, Skelton2014} covers roughly three-quarters (625 arcmin$^{2}$) of the area imaged by the CANDELS ultra-deep survey fields \citep{Grogin2011, Koekemoer2011}. The WFC3/G141 grism is the primary spectral element in the survey, and the G141 grism's wavelength coverage (1.11 $\mu$m $\leqslant \lambda \leqslant$ 1.67 $\mu$m) allows for the detection of H$\alpha$ emission lines for sources at $0.7< z <1.5$. We also used H$\beta$ emission line fluxes from these catalogs when reported. The grism spectroscopic data allows us not only to measure the spectroscopic redshifts (reported in the catalog) but also to use the measured emission line fluxes (after applying dust correction) to determine the star-formation rate of our sources and the underlying attenuation, as discussed in Section 3. The spectroscopic redshifts are particularly used to constrain the multi-band SED fits. Figure \ref{redshift} shows the grism spectroscopic redshift distribution for all the sources in the 3D-HST catalog with ancillary {\it Herschel} observations along with distribution of sources detected in the H$\alpha$ as one of the most prominent nebular emission lines.  

We further took advantage of the {\it HST}/ACS and WFC3 photometry reported for the detected sources in the 3D-HST catalog for the construction of the multi-band SEDs. This included the 3D-HST reported fluxes in near-UV filters (CFHT u; \cite{Erben2009}, KPNO 4m/Mosaic U; \cite{Capak04}, and VLT/VIMOS U; \cite{Nonino09}), optical filters (ACS F435W, F606W, F775W, F814W and F850LP) and near-IR filters (WFC3 F125W, F140W and F160W). The use of these observations provided a fairly comprehensive wavelength coverage from the near-UV to the near-IR, allowing us to constrain the stellar emissions. Note that not all of the listed HST filters are available in all five of the CANDELS fields.

\subsection{CANDELS Photometry}

The Cosmic Assembly Near-infrared Deep Extragalactic Legacy Survey (CANDELS; \citealp{Grogin2011, Koekemoer2011}) covers approximately 800 arcmin$^2$ of the sky over five main extragalactic fields \citep{Grogin2011}. In this work, we used {\it Spitzer}/IRAC 3.6\,$\mu$m, 4.5\,$\mu$m, 5.8\,$\mu$m and 8.0\,$\mu$m observations reported by CANDELS \citep{Galametz2013,Guo2013,Nayyeri2017b,Stefanon2017} in the five fields in addition to the optical/near-infrared data discussed above. These observations employ TFIT {\bf \citep{Laidler07}} which is a robust photometry measurement algorithm using prior information -- such as position and light distribution profiles -- from high-resolution images, to estimate the photometry in a low resolution band. This is crucial for measuring the {\it Spitzer}/IRAC photometry given the PSF is much wider than the high resolution {\it HST} observations and is essential in constraining the stellar mass of higher redshift sources in our sample.

\subsection{Final sample}

To construct our final sample, as discussed above, we used the fourth data release of the HerMES survey with coordinates matched to the 24\,$\mu$m positions and including the 250\,$\mu$m, 350\,$\mu$m and 500\,$\mu$m observations. These sources were matched with the 3D-HST catalogs also with MIPS 24\,$\mu$m positions using a matching radius of 2$^{\prime\prime}$ in the five CANDELS fields of GOODS-S, GOODS-N, COSMOS, AEGIS and UDS. Our choice of matching radius is more conservative than the MIPS 24\,$\mu$m PSF FWHM (at $\sim 6^{\prime\prime}$) and is more consistent with {\it HST} and {\it Spitzer}/IRAC observations. However, we find that the number of matched sources is not very sensitive to this choice and changes minimally upon choosing a $6^{\prime\prime}$ matching radius. We find 6018 matched sources between the 3D-HST and {\it Herschel} HerMES catalogs out of which 1147 have reported H$\alpha$ line fluxes. These sources form the final sample used in this study. {\bf Table \ref{matches}} summarizes the result of our matching in each of the five fields. The resulting catalog was cross-matched with the CANDELS photometric catalog \citep{Galametz2013,Guo2013,Nayyeri2017b,Stefanon2017} to extract the {\it Spitzer}/IRAC photometry in the four bands at 3.6, 4.5, 5.8 and 8.0\,$\mu$m. The final catalog consists of 1147 sources with multi-band photometric observations from the optical to the infrared and reported spectroscopic observations of the H$\alpha$ and H$\beta$ (when available) nebular emission lines. 

\section{Physical properties}

\subsection{Physical Properties from Multi-band SEDs}

We fit observed SEDs of our sources with the {\sc magphys} package (Multi-wavelength Analysis of Galaxy Physical Properties; \cite{daCunha08}), which compares the observed values of flux density to a library of model SEDs at the same redshift, for the entire SED -- from the UV to the far-IR range. We ran {\sc magphys} in the default mode, wherein the library of reference galaxy spectra are generated using Bruzual \& Charlot's \citeyearpar{BC03} stellar population synthesis models in conjunction with Charlot \& Fall's \citeyearpar{CF2000} dust attenuation libraries, and dust emission models from da Cunha et al. \citeyearpar{daCunha08}. 

The \cite{Bruzual2007} version of the Bruzual \& Charlot (2003) code was used to predict the spectral evolution of stellar populations at wavelengths ranging from 91 \AA{} to 160 $\mu$m, aged from $1 \times 10^5$ to $2 \times 10^10$ yr. In these models, timescales of random bursts of star formation are distributed uniformly between $3 \times 10^7$ and $3 \times 10^8$ yr, galaxy ages between 0.1 and 13.5 Gyr, and stellar metallicities between 0.02 and 2 Z$_{\odot}$ (solar metallicity). The probability density function of star formation timescale is nearly uniform over 0 to 0.6 Gyr$^{-1}$ and drops around 1 Gyr$^{-1}$. For each galaxy, both the dust-attenuated and dust-free spectra are calculated, the former using the angle-averaged two-component dust model of Charlot \& Fall (2000), which accounts for the birth of stars in dense molecular clouds which dissipate in timescales of $10^7$ yr. The dust attenuation model thus treats dust attenuation from both the natal birth clouds and the interstellar medium (ISM) and that from just the ISM separately. The da Cunha (2008) dust emission models combine SEDs of the power reradiated by dust in the IR and from dust emitted by the ambient ISM. For the birth clouds, three components contribute to the dust emission: polycylic aromatic hydrocarbons (PAHs), a mid-IR continuum consisting of hot grains at 130-250 K, and warm dust in thermal equilibrium, where the equilibrium temperature is distributed between 30 K and 60 K.

\begin{figure*}[!]
\centering
\includegraphics[trim=1cm 0cm 0cm 0.5cm, scale=0.9]{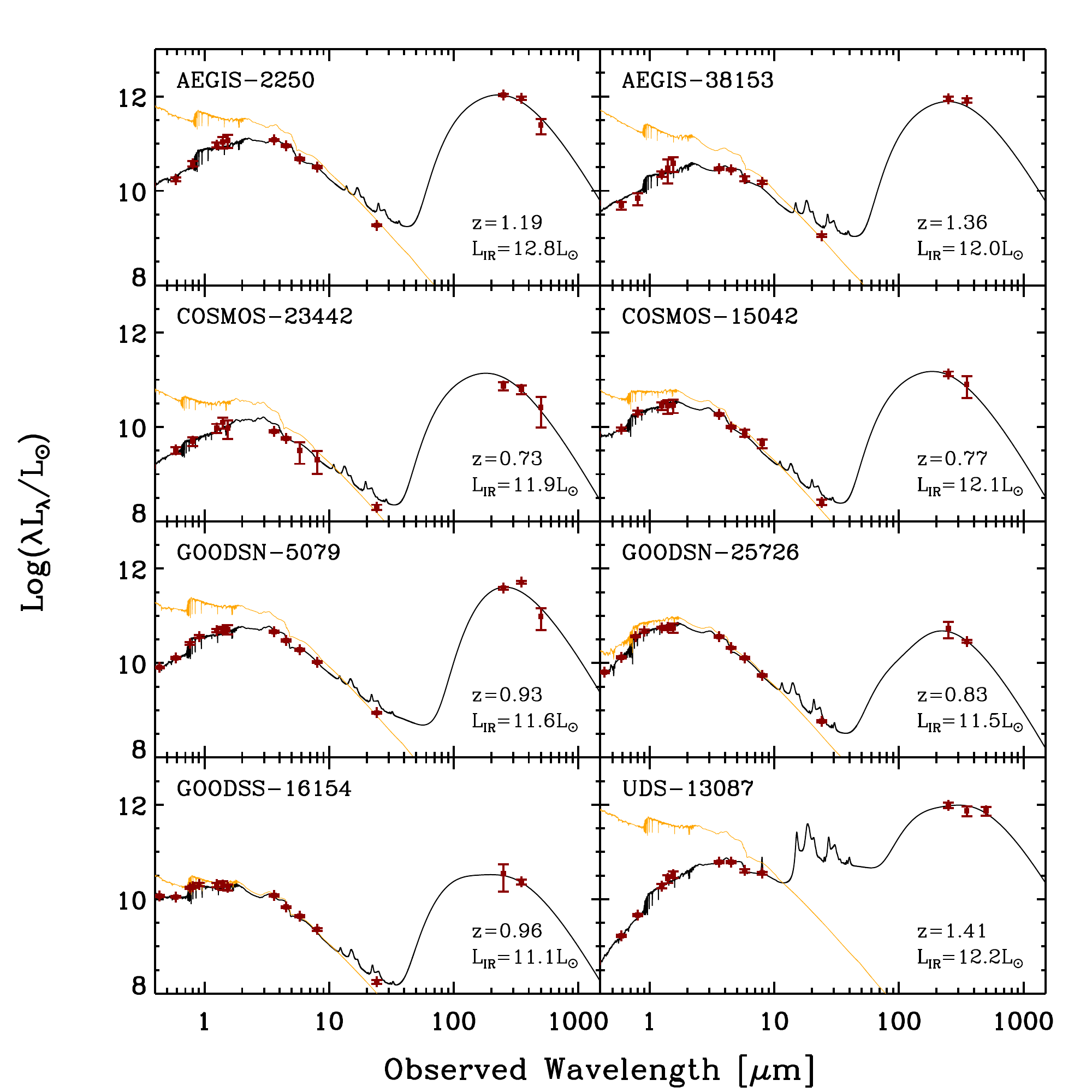}
\caption{Best-fit model SEDs for eight representative sources in the five main fields observed with 3D-HST. The SEDs are constructed using {\sc magphys} \citep{daCunha08}, with optical, near-infrared and infrared observations from {\it HST}, {\it Spitzer} and {\it Herschel} as outlined in the text. In each panel, the best-fit SED model is plotted in black while the intrinsic model without dust attenuation is plotted in orange. For each SED modeling we fix the redshift to the spectroscopic value, as determined by the 3D-HST and is also reported in each panel. The infrared luminosities (rest-frame $\rm 8-1000\,\mu m$) reported in each panel are measured with {\sc magphys} from the modified black-body radiation observed by {\it Herschel}.}
\label{seds}
\end{figure*}

For our SED fitting, we used 13 photometric bands for UDS sources, 14 for AEGIS and COSMOS, 16 for GOODS-N and 17 for GOODS-S. These fits used a near-UV filter ($\sim 0.37\,\mu$m), at least 5 optical filters ($0.5-1.5\,\mu$m), 4 mid-IR filters ($3.6-8\,\mu$m), 3 far-IR filters (250, 350, and 500\,$\mu$m), and with the exception of the UDS sources, the MIPS 24\,$\mu$m mid-IR filter. Thus, we had a comprehensive coverage of wavelengths for fitting our observed SEDs to models at the same redshift. Figure \ref{seds} shows the {\sc magphys} output SEDs for 8 representative sources in the five fields. About half of the fits were very good ($\chi^2 \leqslant 3$), 20\% had $\chi^2 \geqslant 10$, and 10\% were poor fits with $\chi^2 \geqslant 20$. 

{\sc magphys} returns several physical parameters, including total infrared luminosity ($\rm L_{IR}$; rest-frame $\rm 8-1000\,\mu m$), stellar masses ($M_\star$), and dust temperature ($T_d$). These are marginalized parameters, based on the likelihood probability distribution functions from the observed photometry. For the uncertainties, we report the 16\% and 84\% intervals from the probability distribution for each measured parameter. Figure \ref{fir-z} shows the measured total infrared luminosities for our sample of {\it Herschel} detected DSFGs as a function of redshift, with the spectroscopic redshifts derived from the 3D-HST matched catalogs. Our measured luminosities mostly lie above the {\it Herschel} detection limit of $S_{250} >10\,{\rm mJy}$ assuming a dust temperature of $\rm T_{d}=35\,K$, and is in agreement with mid-infrared and millimeter selected samples of dusty galaxies \citep{Casey2012b}. The scatter and deviations from the line is associated with the variations in the infrared SEDs of galaxies due to variations in the intrinsic dust temperatures and opacities.

\subsection{The UV and Far-IR SFRs}

{\sc magphys} measured total infrared luminosities, $\rm L_{IR}$, could be directly converted to total infrared star-formation rates, SFR(IR), using $\rm SFR[M_{\odot}\,yr^{-1}]=10^{-10} \times L_{IR}[L_{\odot}]$ conversion assuming a Chabrier IMF \citep{Chabrier2003}. We further used the rest-frame UV luminosity, $\rm L_{UV}$, measured by {\sc magphys} at 1600\AA, to estimate the UV star-formation rate, SFR(UV), using the Kennicutt's prescriptions \citep{Kennicutt1998}. We added up contributions from SFR(IR) and SFR(UV), derived from these luminosities, to estimate SFR(IR+UV), supposedly the true SFR indicator and compare this with SFR measurements from the {\it HST} grism observations.

\subsection{The H$\alpha$ SFR}

The SFR(H$\alpha$) can be measured directly from the reported H$\alpha$ flux densities assuming a given calibration \citep{Kennicutt1998}. The {\it HST}/WFC3 grism observations lack the resolution necessary to separate the H$\alpha$ from adjacent [N\,II] emission at $\lambda$6548 and $\lambda$6583. Hence, the reported H$\alpha$ flux densities include contributions from both emission lines. Furthermore, light from the H$\alpha$ (and H$\beta$) line includes contributions from atmospheric absorption lines from stars within the galaxy \citep{Dominguez2013}, which needs to be corrected. Finally, the observed H$\alpha$ flux density needs to be corrected for dust attenuation before it can be scaled to SFR(H$\alpha$). At 6563 \AA, H$\alpha$ is less susceptible to dust attenuation than the SFR(UV) discussed above. However, this correction needs to be taken into account for a correct measurement of the nebular SFR \citep{Erb2006}.
 
To correct the reported H$\alpha$ flux density for [N\,II] contamination mentioned above, we used a prescription based on equivalent widths (EWs) of the H$\alpha$ line outlined in \citet{Sobral12}. We used the polynomial relation between $\log$([N\,II]/H$\alpha$) and $\log$\big[EW$_{0}$([N\,II]+H$\alpha$)\big] as outlined in \citet{Sobral12} to correct the [N\,II] contribution.

To account for stellar absorption, we correct the reported emission-line measurements assuming the same equivalent width for H$\alpha$ and H$\beta$ fluxes \citep{Osterbrock1989}. As outlined in \citet{RosaGonzalez2002}, we use the observed ratio between H$\alpha$ and H$\beta$ emission line fluxes:

\begin{equation}
\text{$\frac{F(H\alpha)}{F(H\beta)} = \frac{F_{+}(H\alpha) - F_{-}(H\alpha)}{F_{+}(H\beta) - F_{-}(H\beta)}$}
\end{equation}

where the + and - subscripts are used to denote the intrinsic fluxes of the emission lines and absorption lines respectively. Furthermore, we assumed an intrinsic Balmer decrement ratio of 2.86 \citep{Osterbrock1989} and used the measured equivalent widths of the nebular emission lines in order to determine the Balmer absorption, $Q$, i.e. the ratio of equivalent width of H$\beta$ in absorption to emission ($EW_{-}(H\beta)/ EW_{+}(H\beta)$) from the following equation:

\begin{equation}
\text{$\frac{F(H\alpha)}{F(H\beta)} = \frac{2.86 \bigg[1-Q \frac{EW_{+}(H\beta)}{EW_{-}(H\alpha)} \bigg] }{1-Q}$}
 \end{equation}

Using these ratios with proportionality relation of fluxes of emission lines to their EWs, we corrected H$\alpha$ and H$\beta$ fluxes for stellar absorption.

To correct the observed H$\alpha$ fluxes for dust attenuation, we applied Calzetti's \citeyearpar{Calzetti2001} prescription. We had 26 sources with S/N $\geqslant$ 3 for both H$\alpha$ and H$\beta$ fluxes, for which we could robustly utilize the Balmer decrement to calculate the dust attenuation of the H$\alpha$ emission line. We computed the Balmer optical depth, the logarithmic ratio of observed to intrinsic Balmer decrement. Finally, assuming a Calzetti reddening curve, we computed the nebular attenuation.

To apply dust correction for sources that didn't have secure detections for both H$\alpha$ and H$\beta$, we used the V-band optical depth of each source, $\tau_{v}$, output by {\sc magphys} SED modeling and applied it to the definition of $\tau$ and $A_{\lambda}$: $A^0_{\lambda} = 1.086 \tau_{\lambda}$, where $A^0_{\lambda}$ is the attenuation at the wavelength $\lambda$. Then, assuming the \citet{Calzetti2000} reddening curve for stellar extinction, we obtained the color excess, $E_{star}(B-V)$. 

Next, we applied a differential f-factor to convert the stellar color excess to a nebular color excess, $E_{neb}(B-V)$. \citet{Calzetti2000} defined this as a differential attenuation between nebular emission lines and the stellar continuum and parameterized this scaling factor as:

\begin{equation}
\text{$E_{star}(B-V) = f \times E_{neb}(B-V)$,}
\end{equation}

and, in a sample of local starbursts, found $f=0.44 \pm 0.03.$ Essentially, this implies that ionized gas is about twice as attenuated than stars. More recently however, in a study of {\it Herschel} sources at $0.7<z<1.5$ in the GOODS-S, \citet{Puglisi16} derived a value of 0.93 for this f-factor. \cite{Puglisi16}'s sample was also selected from {\it Herschel} and 3D-HST-matched galaxies and contains galaxies at similar redshifts to ours, in contrast to the nearby starbursts in \cite{Calzetti2000}. They also obtained physical parameters from {\sc magphys} SED-fitting. Due to the similarities in our samples and methodologies, we used the value reported in \cite{Puglisi16} for our own calibration between nebular and stellar dust extinction.

\begin{figure}
\begin{center}
\includegraphics[trim=2cm 0cm 0cm 0cm, scale=0.45]{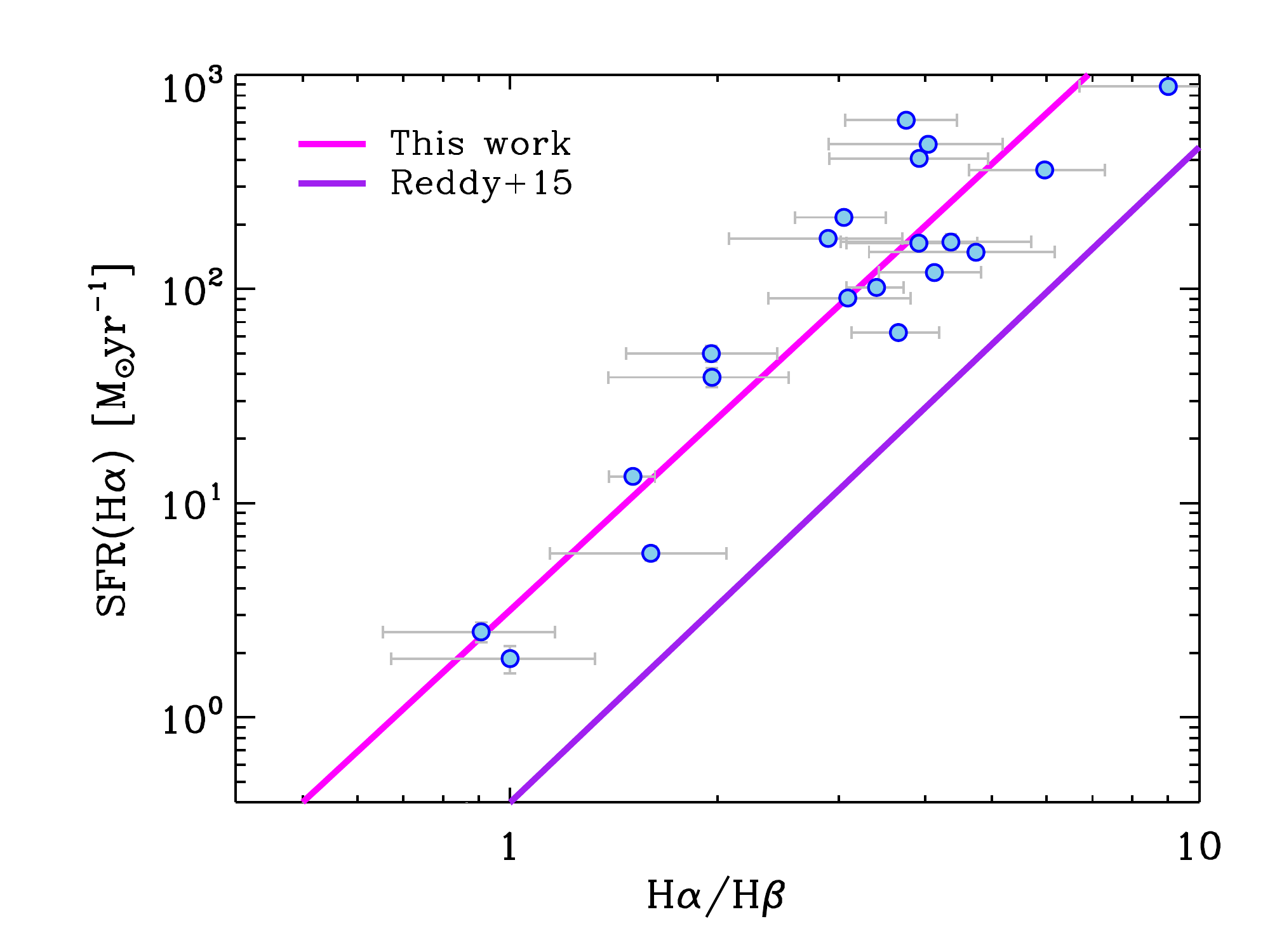}
\caption{The SFR(H$\alpha$) as measured from the corrected H$\alpha$ flux density as a function of the Balmer decrement for the sub-sample of dusty galaxies for which both lines were detected with a signal-to-noise ratio of at least 3. The increased attenuation as a function of star-formation activity is observed in dusty star-forming galaxies, similar to previous studies of star-forming systems at $z\sim1-2$ \citep{Reddy15}.}
\label{sfr-hb}
\end{center}
\end{figure}

\begin{figure}
\begin{center}
\includegraphics[trim=2cm 0cm 0cm 0cm, scale=0.45]{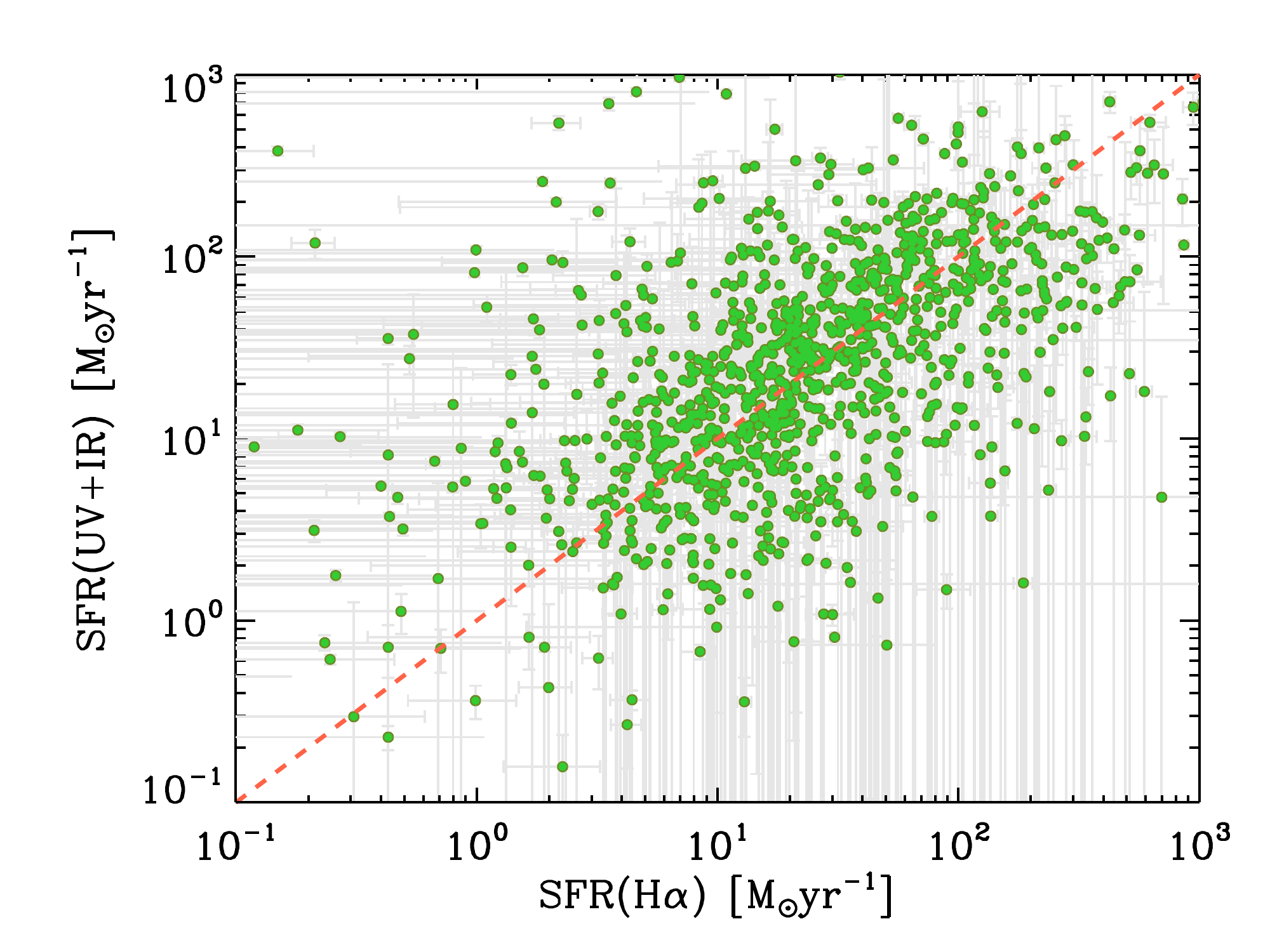}
\caption{SFR(IR+UV) derived from UV and total IR luminosities measured by {\sc magphys} as described in the text versus SFR(H$\alpha$) from grism observations. The measured H$\alpha$ line fluxes are corrected for attenuation, stellar absorption and contribution from adjacent [N\,II] lines in the grism.}
\label{sfr}
\end{center}
\end{figure}

Figure \ref{sfr} shows the SFR(H$\alpha$) measured from the corrected H$\alpha$ flux density, as discussed above, versus the SFR(IR+UV). The general agreement between the two SFR indicators validates our choice of 0.93 for the f-factor. The f-factor of near unity to match the SFR (H$\alpha$) to SFR(IR+UV) means that differential dust extinction is lower in the higher redshift universe than in the local one. In Section 4.1, we study the dust-dependence of SFR(H$\alpha$) and obtain a relation between SFR(H$\alpha$) and the balmer decrement (ratio of H$\alpha$ to H$\beta$).

\subsection{AGN contamination}

The multi-band SED modeling results and observations of the rest-frame optical lines such as the H$\alpha$ and continuum emission in the UV and IR can potentially suffer from ambiguities in the presence of an AGN component \citep{Shipley16}. To search for possible AGN-host galaxies, we matched our combined final catalog to the X-ray catalogs from Chandra \citep{Elvis09}. Nearly 10\% of the sources in our final sample were matched to a counterpart in the X-ray catalogs. For these sources, we fit their SEDs with an AGN template using the {\sc sed3fit} code \citep{Berta13}. {\sc sed3fit} fits galaxy SEDs by combining a stellar emission, dust emission and possible dusty torus/AGN component, using the {\sc magphys} IR dust library and Bruzual \& Charlot \citeyearpar{BC03} population synthesis models. We found virtually no difference in the measured physical properties of our X-ray detected sources with these measurements, leading us to believe AGN activity did not significantly affect our observations and consequently does not alter the main scientific conclusions of this work.

\section{Discussion}

\subsection{SFR-dependence of Dust attenuation}

The correlation between total dust attenuation and star-formation rates has been observed at both low and high redshifts, and can be explained by the ISM of the galaxy being enriched in dust with increasing SFR activity \citep{Reddy15}. To test this dependence, we observed the relationship of (H$\alpha$/H$\beta$) -- the Balmer decrement -- with SFR(H$\alpha$), for the 26 sources for which we had secure (with $\rm S/N \geqslant 3$) H$\alpha$ and H$\beta$ observations. For this sub-sample we used the observed Balmer decrement directly to correct the measured SFR(H$\alpha$) for dust attenuation. Figure \ref{sfr-hb} shows the relation between measured star-formation activity and observed Balmer decrement. This yields a best-fit power-law relation:

\begin{equation} \label{eqhb}
\text{$\log$\big(SFR(H$\alpha$)\big) = 2.98 $\times$ $\log$\big((H$\alpha$/H$\beta$)$_{\rm obs}$$\big) + 0.50$}
\end{equation}

We used this relation to estimate SFR(H$\alpha$) for the rest of the sources that had H$\alpha$ and H$\beta$ reported. This enabled us to compare a lower limit for SFR(H$\alpha$) computed using Balmer decrements with those obtained from Calzetti optical depth and nebular color excess derived from the differential f-factor discussed above. We note, however, that eq. (\ref{eqhb}) was obtained from a very small fraction of the total sample and may not be applicable to the overall sample of DSFGs. Figure \ref{sfr-hb} further shows the observed relation for a sample of star-forming galaxies from Keck near-IR spectroscopic observations \citep{Reddy15}. The {\it Herschel} detected dusty galaxies in our sample sit above the relation of normal star-forming galaxies, indicating greater star-formation activity for any given balmer decrement.

\subsection{The IRX-$\beta$ Relation}

To compare dust extinction from the UV to that of the IR, we studied the IRX-$\beta$ relationship. Energy conservation demands that the photons absorbed in the UV and optical should be re-radiated in the infrared. Thus, for a measure of the total dust attenuation, we computed the IRX (Infrared Excess Ratio), defined as $L_{IR}/L_{UV}$ \citep{Narayanan2018, Reddy2018}. Assuming each deviation from the constant-shaped UV spectra of star-forming galaxies is produced by dust, we computed the UV slope, $\beta$, using \citet{Meurer99} calibration, based on local starburst galaxies, and the continuum attenuation at rest-frame wavelength (1600\,\AA) following:

\begin{equation}
\text{$A_{1600}=4.43+1.99 \beta$}
\end{equation}

\begin{figure}
\begin{center}
\includegraphics[trim=2cm 0cm 0cm 0cm, scale=0.45]{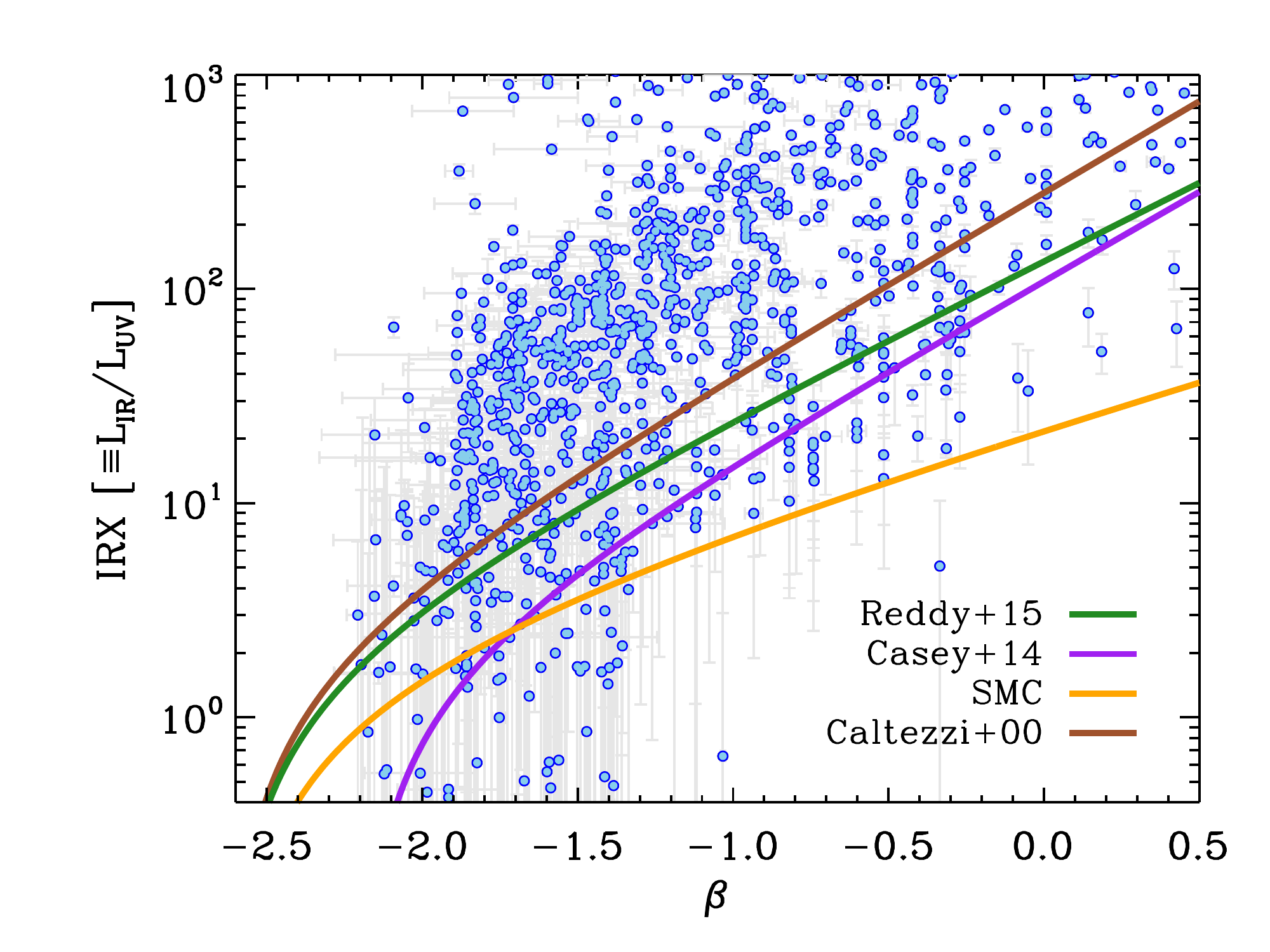}
\caption{The IRX-$\beta$ relation. Most Dust-obscured galaxies appear to be skewed toward bluer colors, i.e. smaller UV slopes. Our sample lies above the expected relations at both low and high redshifts, including for local star-busts \citep{Meurer99}, MOSFIRE galaxies \citep{Reddy15}, and DSFGs at up to $z\sim3.5$ \citep{Casey2014}.}
\label{irx-beta}
\end{center}
\end{figure}

\citet{Meurer99}'s sample was among the first to produce a strong correlation between UV spectral slope -- essentially a color -- and IRX, indicating that dust absorption is directly related to UV reddening. Later studies confirmed the relation for both local \citep{Seibert05, dePaz07} and high redshift \citep{Reddy2010, Reddy2012, Casey2014} galaxies. \cite{Casey2014} also found that DSFGs with high star formation rates lie significantly above the expected IRX-$\beta$ relationship. The IRX-$\beta$ relationship is shown in Figure \ref{irx-beta}. The trend observed is similar to, but above, that expected of starburst galaxies. This is expected for {\it Herschel} detected galaxies that re-emit most of their UV light in infrared, as is the case for DSFGs. 

\subsection{The $SFR-M_{\star}$ relation}

The star-formation rate (SFR) and stellar mass ($M_\star$) of star-forming galaxies (SFGs) are directly correlated across various cosmic epochs. This so-called main sequence of star formation (MS; \citealp{Noeske2007}), has been used extensively to test models of galaxy formation and evolution at different redshifts \citep[e.g.][]{Brinchmann04, Elbaz07, Guo15, Shivaei15}. Observations, in general, show that star-formation activity increases as a function of stellar mass for SFGs and that this a strong function of redshift \citep{Elbaz2011, Pannella2015}. At higher redshifts galaxies of similar stellar mass tend to be more star-forming as demonstrated from direct observation of star-formation activity \citep{Elbaz2011, Nelson2016} and from observations of molecular gas reservoirs \citep{Genzel2015, Scoville2016}.

\begin{figure}
\begin{center}
\includegraphics[trim=2cm 0cm 0cm 0cm, scale=0.45]{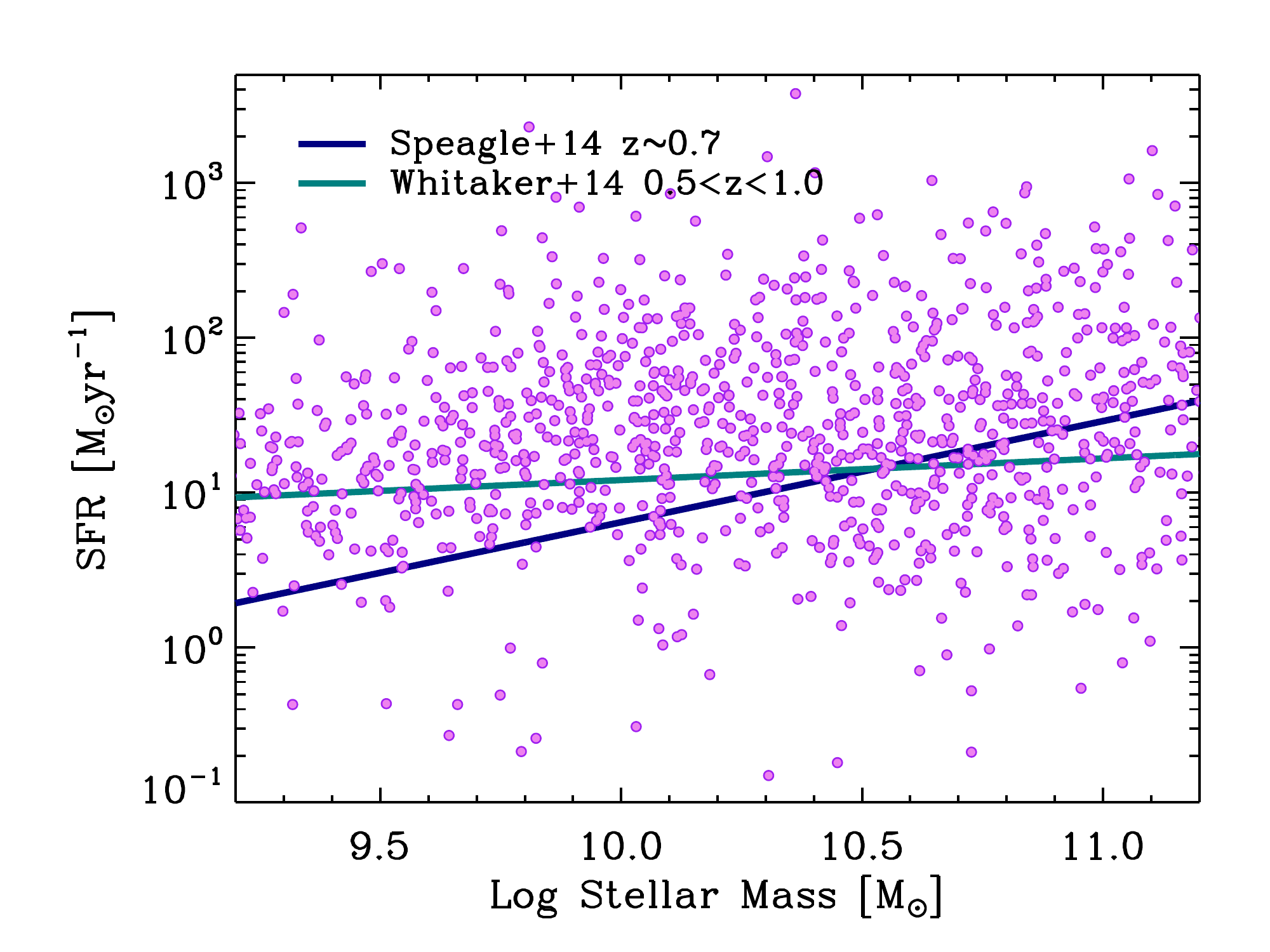}
\caption{The main sequence of star-formation for our sample. The SFR reported are measured from the corrected H$\alpha$ flux densities assuming a \citet{Kennicutt1998} conversion, as discussed in the text. The stellar masses are measured from the multi-band SED modeling using {\sc magphys} including observations in the near-infrared from {\it HST}/WFC3 and {\it Spitzer}/IRAC. The near-infrared observations of these dusty systems was crucial in constraining the stellar masses. We see that the population of DSFGs at $0.5<z<1$ show elevated SFR at any given stellar mass compared to the main sequence of star-formation as observed for normal star-forming galaxies \citep{Speagle14, Whitaker14}.}
\label{ms}
\end{center}
\end{figure}

Figure \ref{ms} shows the SFR versus $M_\star$ for our sample of {\it Herschel} selected dusty star-forming galaxies. The DSFGs sit above the MS relation at similar redshifts compared to that of normal star-forming systems \citep{Speagle14, Whitaker14}. This is consistent with recent studies of sub-millimeter bright galaxies pointing towards enhanced star-formation activity and/or larger molecular gas reservoirs accesible to these systems \citep{Scoville2016}. The scatter is mostly associated with diverse star-formation histories and uncertainties in dust correction assumptions and stellar mass derivations from multi-band SED modeling. 

\section{Summary and conclusion}

We presented a multi-wavelength study of 1147 spectroscopically-identified dusty star forming galaxies at $0.49<z<2.24$ in the five CANDELS fields. We constructed SEDs for each of these sources, from the UV to the far-IR, using photometry from {\it HST} and {\it Herschel} and spectroscopic redshifts measured from nebular emission lines. Physical properties of these galaxies, such as stellar mass, $M_\star$, IR luminosity, $\rm L_{IR}$, and V-band optical depth, $\tau_{V}$, were obtained by comparing observed SEDs with SEDs at the same redshift predicted by Bruzual \& Charlot's \citeyearpar{BC03} population synthesis models. We used the 3D-HST H$\alpha$ grism line measurements to infer SFR from nebular lines and IR and UV luminosities to infer SFR from the stellar continuum, having accounted for the dust attenuation for all these quantities. 

Our main conclusions are as follows:

\begin{itemize}
 
\item The differential f-factor of 0.93 used by \cite{Puglisi16} used to relate nebular dust attenuation to that of the stellar continuum gave us good agreement between the H$\alpha$ and IR+UV derived SFRs. This implies that the extinction of ionized gas is close to that of the stellar continuum, running counter to the generally accepted idea that nebular emission lines are about twice as attenuated as the stellar continuum.
\item We observed a correlation between IRX and UV spectral slope, $\beta$, and our galaxies sit well above the established relation for normal star-forming galaxies. Dusty galaxies are observed to have preferentially bluer colors.
\item We observe that star formation rates increase with increased dust attenuation, as is typical of DSFGs.
\item We also studied the star formation main sequence relationship between SFR and $M_\star$ of these high-$z$ galaxies. Our distribution, skewed toward higher masses, shows SFRs elevated relative to the well-established SFR-M$_\star$ relation, and supports the idea that luminous galaxies have higher star formation rates. 
\end{itemize}

\section*{Acknowledgement}
F.H. was financially supported by the Summer Experience Award from Reed College International Student Services. Financial support for this work was provided by NSF through AST-1313319 for H.N. and A.C. UCI group also acknowledges support from HST-GO-14083.002-A, HST-GO-13718.002-A and NASA NNX16AF38G grants.

\bibliographystyle{apj}
\bibliography{reference}

\end{document}